GAIA MICCA LONGO[*], VINCENZO LAPORTA[**], SAVINO LONGO[*,**]

[*] DIPARTIMENTO DI CHIMICA, UNIVERSITÀ DEGLI STUDI DI BARI ALDO MORO – VIA ORABONA 4 – 70125 BARI, ITALY
[**] ISTITUTO PER LA SCIENZA E TECNOLOGIA DEI PLASMI, CONSIGLIO NAZIONALE DELLE RICERCHE, VIA AMENDOLA 122/D, 70125 BARI, ITALY

# New insights on prebiotic chemistry from plasma kinetics

**Summary** – The famous Miller-Urey experiment, which provides essential information on the prebiotic synthesis of the molecules of life, still has many obscure points. In this paper, we want to suggest a way of possible future progress, which consists in framing the experience of Miller and Urey in the context of the kinetics of ionized gas, or plasma. In this framework, extremely effective and versatile theoretical tools, based on quantum mechanics and chemical kinetics, make it possible to look, in a new way, at the elementary processes that lead to the formation of excited species and ions, at the base of the cascade of subsequent reactions.

*Introduction*

Earth's chemical evolution and the origin of life represent one of the most important issues in astrochemistry and astrobiology. Two different theories [1] have been proposed regarding the emergence of life on Earth: (i) exogeneous delivery and (ii) endogenous synthesis. According to the first theory, prebiotic molecules have reached the Earth surface by means of comets, meteors and asteroids. The so-called delivery, *i.e.* the actual transport of molecule from space to Earth, represents the key point of the Panspermia theory [2,3]. It was demonstrated that carbonate and sulfate micrometeoroids (especially, calcite and anhydrite) may act as good prebiotic molecule carriers [4-6], in view of the well-known association between carbonates/sulfates and organic matter presence [7-9]. As to the endogenous synthesis, the Miller-Urey experiment simulated the early Earth chemical condition than might have helped the organic molecule formation [10-12].

In this work, in the context of the endogenous synthesis, we want to show how recent developments in the theoretical modeling of chemical reactions and transport phenomena in weakly ionized gases and in thermodynamic non-equilibrium conditions can help to put the studies on the synthesis of prebiotic molecules in the context of a primordial atmosphere, carried out in the 50s and 60s, under a new light.

The composition and depth of the primordial atmosphere are the subject of a lively debate in planetology. Tracing back to Oparin [13] and Haldane [14], most scientists believe that, at the very least, the primordial atmosphere was enriched with reducing low molar mass components, like methane or ammonia or even hydrogen. There is no reason to assume that an early Earth atmosphere (with a mixture of components of such very different molar mass, like nitrogen and hydrogen) should have uniform composition in its lower and upper regions: most probably, the early atmosphere was enriched in low mass components at high altitudes.

The topic of primordial atmosphere as source of prebiotic matter dates back to the Miller-Urey experiments, during the 1950s [10-12], partially modeled on Oparin's hypothesis. They speculated that primordial atmospheres of terrestrial planets are filled with methane, ammonia, and hydrogen; moreover, they theorized that the primordial clouds provide the necessary energy to activate small molecules and link them into bigger molecules, in order to synthesize the building blocks of life.

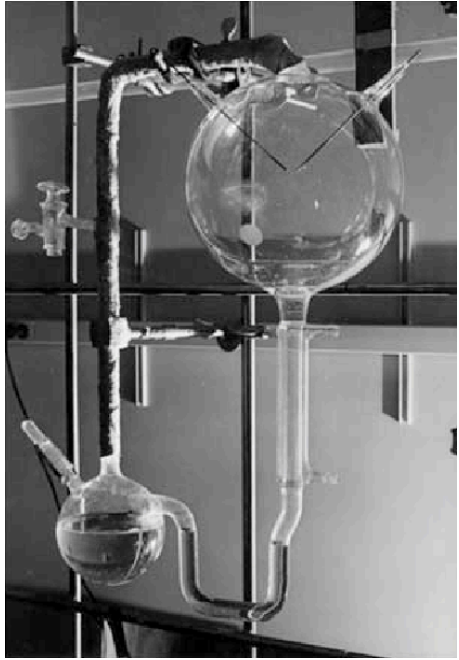

Figure 1. The Miller-Urey apparatus. The lower flask contains the primordial ocean model, while, through the gas inlet/outlet on the right, the rest of the apparatus is filled with the strongly reducing primordial atmosphere, based on the knowledges of Miller's times. The third neck of the lower flask is used to sample the liquid for the analysis. The higher balloon displays the spark gap. Photo adapted from [15]

The experimental apparatus (figure 1) consisted of two glass balloons: the lower one was partially filled with liquid water (primordial ocean); a gas mixture composed by methane, ammonia and hydrogen (primordial atmosphere) filled the rest of the inside. In this primordial gas, Miller placed two electrodes to simulate lightning-like electric discharge. Therefore, Miller and Urey recreated the disequilibrium chemistry by applying electrical discharge to the gases that probably formed a highly reducing primordial atmosphere. The two spheres in the apparatus were connected in a hydrological cycle, that allowed mimic rain after the electrical discharge. The analyses demonstrated that, after the experiment, the primordial ocean was rich in carbon compounds, which contained significant amounts of amino acids. Miller and Urey demonstrated that organic molecules might be formed spontaneously from inorganic components, in such primordial reducing environments.

With respect to the studies carried out in the few years following the Miller experiment, considerable progress is available, in terms of theoretical and methodological understanding. Moreover, the hypotheses regarding the composition of the primordial terrestrial atmosphere have changed considerably.

Since the 1980s, geoscientists have been questioning about Miller's hypothesis of highly reducing primordial atmosphere. Many experiments were carried out using CO and $CO_2$ atmospheres [16,17]; however, the synthesis of organic compounds by the action of electric discharges on neutral gas mixtures is much less efficient, in this case. On the other hand, results presented by Cleaves *et al.* [18] demonstrate that neutral atmospheres can produce amino acids in much higher amount than previously thought.

Hypotheses regarding the chemical composition of the Earth's earliest atmosphere are numerous, and no definite answer can be found in the literature. Moreover, it is believed that there were many atmospheres over the vastness of Earth's geological time [19]. Geologically based arguments, that treat the atmosphere as outgassed from solid Earth, indicate that the Earth's primitive atmosphere was composed mostly of $H_2O$, $CO_2$ and $N_2$, with traces of CO and $H_2$ [20-22]. A complete and exhaustive review about the origin and conditions of the Earth's earliest atmospheres can be found in the work by Zahnle *et al.* [23], with a particular attention to the issue concerning the origin of life, which is set during the so-called Hadean Earth (~ 4.6 - 4 Gys ago).

Under the hypotheses that the primordial Earth's atmosphere consisted predominantly of CO and $CO_2$, the chemical channels previously considered must be reformulated and submitted to new studies, bearing in mind that other components, even minority ones, must play a role in subsequent chemical reactions.

Although the theoretical studies on this system are advanced [24-26], one aspect of these studies is still rather primitive, namely the role of electrons in the plasma (a short word for an ionized gas introduced by I. Langmuir) in producing the first excited and radical species and ions, which later enter several chemical channels to form prebiotic molecules. Studies of chemical kinetics performed in the case of gas discharges with chemical composition enormously simpler than the one used in Miller-type experiences have shown that ionization excitation processes have an extremely complex and interesting kinetics, due to the non-equilibrium conditions. In a plasma, the electrons have a much greater kinetic energy than those of the atomic and molecular neutral species, and the distributions of speed and energy are not trivial. These studies show that the efficiency of radical formation can be much greater than those observed in the much more phenomenological studies devoted to Miller's experience so far.

*Kinetics of $CO_2$ activation*

Compared to previous studies, the main reaction channel becomes the formation of reactive intermediates with the help of the energy available in the plasma starting from $CO_2$.

An important aspect of Miller-related chemistry is how to avoid all carbon being chemically trapped into $CO_2$, which is a remarkably stable molecule: it appears that some processes, under the conditions of a primordial atmosphere, must be able to activate chemically this molecule with the help of physical energy in the form of, *e.g.*, kinetic energy of the free electrons.

In this respect and in recent years, great progress has been made in the theoretical understanding of the chemical plasma activation of $CO_2$, due to the importance that this process plays in the $CO_2$ recycling study for the mitigation of the climatic consequences of $CO_2$ accumulation in the atmosphere [27,28].

Since the most effective dissociation path of $CO_2$ is through the asymmetric stretching mode, it is necessary to distinguish the 22 vibrational levels of this mode. Electron-impact processes, the so-called 'eV processes', however, can only pump energy into the lowest vibrational levels, say the first 4 or 5:

$$e + CO_2(v) \rightarrow e + CO_2(v+1).$$

Subsequently, $CO_2$ molecules reach the dissociation limit with the help of a different process, named $VV_1$, where a low excited $CO_2$ molecule interacts with a more excited one by pushing it to higher excitation:

$$CO_2(1) + CO_2(v) \rightarrow CO_2(0) + CO_2(v+1).$$

This mechanism, proposed by russian scientists about 40 years ago [29], works like a "conveyor belt", where molecules move up the internal energy ladder until they reach the dissociation threshold, *i.e.*:

$$CO_2(1) + CO_2(21) \rightarrow CO_2(0) + CO + O.$$

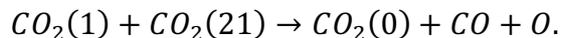

This process is extremely effective in increasing the efficiency of activation, since the direct dissociation channel with electronic states as intermediates has an energetic cost in the order of 10 eV.

In the last few years, many researchers aimed to a better understanding of this mechanism using computer simulations. A Dutch-Italian collaboration developed a scheme based on the characterization of the upwards flux in energy space, $J$ [30-32]. $J$ is a functional of the function $f(\varepsilon)$, which is the distribution of molecules according to their internal energy $\varepsilon$. The mathematics of diffusion, namely, the continuity equation of the flux on the internal energy space:

$$\frac{\partial f}{\partial t} = -\frac{\partial J}{\partial \varepsilon},$$

demonstrates that at the steady-state $J$ is a constant. Therefore, $J$ is the dissociation frequency per molecule, since $\varepsilon$ can be assumed to be equal to the dissociation energy.
A study of the possibility of a "conveyor belt" regime, where upward drift in energy dominates the diffusion, or spreading, of energy conditions, was in order. The mathematics of diffusion processes shows that $f$ can be easily calculated by solving the Fokker-Planck equation based on rate coefficients of vibrational processes [31]. But, since $J$ needs $f$ to be calculated, the process is iterated until self-consistency is attained, and the problem is solved. At this point, we observe the figure 2, which is based on calculations of the described type: in all cases on the x-axis, we set the internal energy, which goes from the value 0, corresponding to the zero-point energy of the quantum oscillator that describes the asymmetric stretching motion, to the dissociation energy. The plotted functions are precisely the internal energy distributions. The two plots differ in the value of $T_{v01}$, the excitation temperature of the 01 transition.

Comparing the plot on the left (lower temperature) with the one on the right (higher temperature) and observing the values of $J$ corresponding to the different curves shown in each of the two figures, two completely different behaviors are observed. While in the case of lower temperature, the function can only be modified with considerable variations of the functional $J$, in the case of high temperature, completely different energy distributions are obtained for very small variations of the functional $J$. What does this mean? On the one hand, the low-temperature case exemplifies that the energy distribution determines the dissociation rate, especially in the high-energy region, which is closer to where the dissociation takes place. On the other hand, in the high temperature case, a regime is realized in which the functional $J$, and therefore the dissociation rate of the molecules, changes very little even in the hypothesis of enormous variations of the high energy part of the distribution.

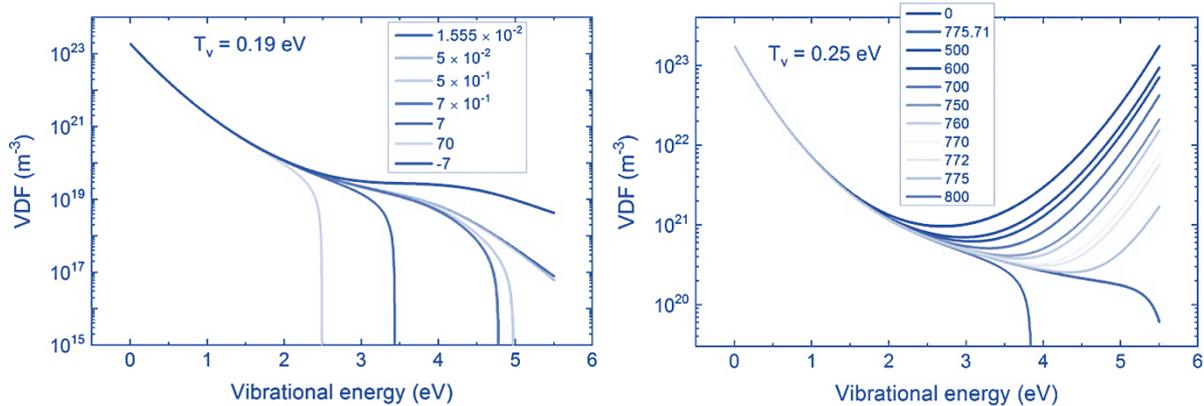

Figure 2. Calculated distribution of internal energy in $CO_2$ molecules based on the functional method $J//f$ described in the text, for two different values of the vibrational temperature $T_v$. On the left: a lower temperature regime where the specific dissociation rate $J$ is strongly determined by the high energy region of the distribution function $f$. $J$ is almost insensitive to $f$, and molecules keep drifting up the energy scale. On the right, the high $T_v$ regime where the shape of the distribution is very sensitive to $J$: correspondingly, the dissociation rate is rather insensitive to high-energy processes. Figure adapted from [30].

This result is very important in a reconsideration of the Urey-Miller experiment with an atmosphere rich in $CO_2$. In such an atmosphere, locally produced plasmas may produce radicals from $CO_2$ with a very high energetic efficiency, which was unforeseeable at Miller's times.

*Elementary processes of CO and $CO^+$*

Many important progresses have been reaching in the ionization efficiency aspect: this parameter is important since most of the molecular formation channels of biological interest is ion channels, starting from the monomers in Miller's experience. In the 80's, computer simulations developed for a better understanding of chemical lasers, in particular of excimer lasers, showed that the preliminary excitation of metastable states produces ionization efficiencies much greater (orders of magnitude) than those estimated neglecting the effect of excitation [32]. For this reason, the recent studies related to rigorous calculations of the excitation cross sections of molecules relevant in the Miller experience in $CO_2$ become important.

Chemically active components, that derive from the chemical plasma activation process of carbon dioxide, are CO and atomic oxygen; in particular, CO is extremely important as a reactive molecule in the synthesis processes in organic chemistry both in neutral form, electronically excited species $CO^*$ or $CO^+$ positive ion.

In recent years, calculations based on rigorous methods of quantum mechanics, and motivated by the great importance of the CO and $CO^+$ species in astrophysics and in aerospace technologies, have led to great improvements in the quantitative understanding of the reactivity of these species in a low-ionized plasma. In particular, *ab-initio* studies based on UK-R-Matrix quantum chemistry code [33], and by the approaches of Multichannel-Quantum-Defect-theory (MQDT) [34] and of Configuration-Interaction (CI) [35], state-resolved cross-sections and rate coefficients have been available in literature. Figure 3 shows the potential energy curves for CO and the corresponding resonant state $CO^-$ as a function of the internuclear distance $R$; the CO wave function $\chi(R)$, corresponding to the vibrational level $v = 10$ and the resonant wave function $\xi(R)$ for an incoming electron energy of 5 eV are also shown. The cross section by electron-impact is calculated from the following formula:

$$\sigma(\epsilon) = \frac{16\,\pi^4 m}{\hbar^2}\frac{k_f}{k_i}\,|\langle\chi(R)|\mathcal{V}(\epsilon,R)|\xi(R)\rangle|^2\,,$$

where $k_i(k_f)$ is the initial (final) momentum of the electron and $\epsilon$ the corresponding energy; $\mathcal{V}(\epsilon,R)$ is the coupling between the neutral and the resonant state. Figures 4 and 5 show the results for the rate coefficients for electron collisions by CO [36,37] molecule obtained in the CI method and for CO$^+$ [38] molecular ion obtained in the MQDT framework, respectively.

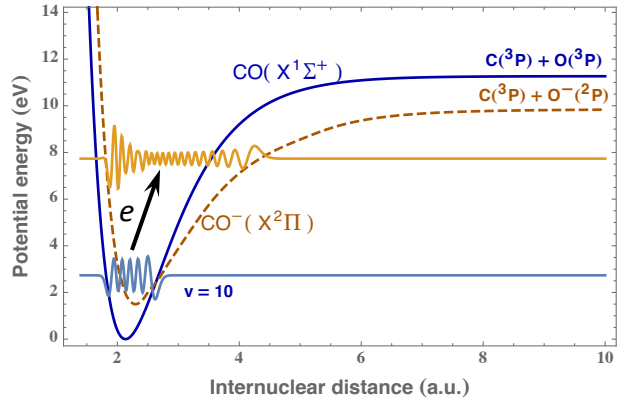

Figure 3. R-Matrix *ab-initio* potential energy curve of CO molecule (blue line) and of the resonant state CO$^-$ (dashed orange line). CO vibrational wave function for v = 10 and the resonant wave function of an electron energy of 5 eV are shown.

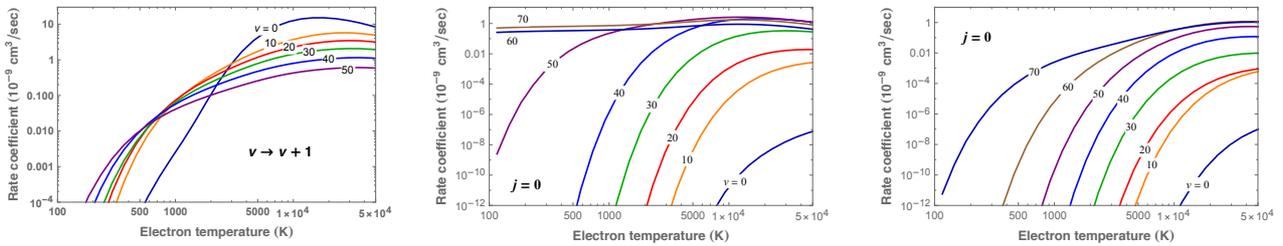

Figure 4. Rate coefficients state-resolved for electron-CO collisions [36,37]. Processes of: Vibrational excitation (left panel); Dissociative attachment (middle panel); Dissociative excitation (right panel).

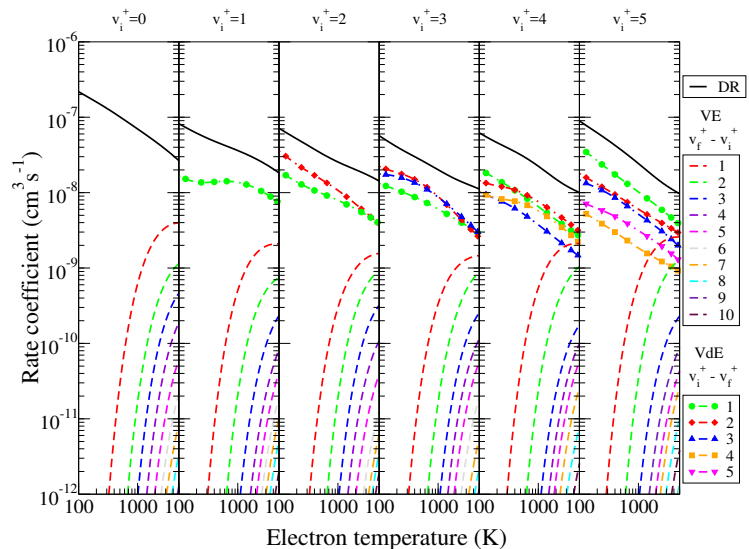

Figure 5. Summary on rate coefficients for electron-$CO^+$ collisions [38] for Dissociative Recombination (DR) and vibrational excitation (VE and VdE) processes.

*Conclusions*

In this short account, we have presented a new perspective on future research in the chemical kinetics of the first active species of the primordial atmosphere and their subsequent role into the formation of the first prebiotic species. The framework of this study is the Urey-Miller experiment, its improvement and understanding, and the clues it provides into the possibility that life machinery arose as a result of an abiotic process. Two new elements of knowledge may produce a synergistic push towards further progresses: the first, the awareness that the primordial atmosphere was not at all the strongly reducing mixture with essentially solar-nebula composition believed in Miller's times. The second is the development of new methods in the context of computer modeling of the kinetics of plasmas, motivated by ecological, astrophysical and aerospace problems. The communication between the two communities of plasma kinetics and astrobiology can therefore help, in the future, to attain a better understanding and new insights on the chemical kinetics of an historical experiment, which has changed our ideas on the genesis of prebiotic molecules on the primordial Earth.